\documentclass[a4paper,12pt]{article}
\usepackage[T1]{fontenc}
\usepackage[utf8]{inputenc}
\usepackage[margin=14mm]{geometry}% <-------- CHANGE HERE for the global margins
\usepackage{mathptmx} % for Times
\usepackage{graphicx}
\usepackage{subfig}
\usepackage{float}
\usepackage{authblk}
\usepackage[superscript,biblabel]{cite}
\usepackage{url}
\usepackage{ulem}
\usepackage{xcolor}

\title{Reduced network extremal ensemble learning (RenEEL) scheme for community detection in complex networks}

\author[1,2]{Jiahao Guo}
\author[1,2]{Pramesh Singh}
\author[1,2,3]{Kevin E. Bassler\thanks{Corresponding Author: bassler@uh.edu}}
\affil[1]{Department of Physics, University of Houston, Houston, Texas 77204, USA.}
\affil[2]{Texas Center for Superconductivity, University of Houston, Houston, Texas 77204, USA.}
\affil[3]{Department of Mathematics, University of Houston, Houston, Texas 77204, USA.}
\date{}
\begin{document}

\maketitle
\begin{abstract}
We introduce an ensemble learning scheme for community detection in complex networks. The scheme uses a Machine Learning algorithmic paradigm we call Extremal Ensemble Learning. It uses iterative extremal updating of an ensemble of network partitions, which can be found by a conventional base algorithm, to find a node partition that maximizes modularity. At each iteration, core groups of nodes that are in the same community in every ensemble partition are identified and used to form a reduced network.
Partitions of the reduced network are then found and used to update the ensemble. The smaller size of the reduced network makes the scheme efficient. We use the scheme to analyze the community structure in a set of commonly studied benchmark networks and find that it outperforms all other known methods for finding the partition with maximum modularity.
\end{abstract}

\section*{Introduction}
Among the most basic and important problems in Network Science is to find the structure within a network~\cite{fortunato2010,newman2004finding}. One way of doing this is to find the community, or modular structure of the nodes. In many real-world networks, the community structure has been found to control much of their dynamical or functional behavior.
Although there are many possible definitions of community~\cite{schaub2017,peel2017}, a commonly used definition assumes that a community is a group of nodes that are more densely connected than what would occur randomly.
This intuitively appealing concept of community can be used to define a metric, called {\it Modularity} $Q$, that quantifies 
the extent to which a partition of the nodes of a network is modular~\cite{newman2004finding}.
The community structure of a given network can then be obtained by finding the partition of the network's nodes that has the maximum modularity $Q_{\max}$. Finding this partition, however, is an NP-hard problem~\cite{brandes2008modularity}. 
It is of considerable interest and importance to develop an algorithm that robustly finds an accurate solution to this optimization problem that completes in polynomial time. 
The accuracy of a solution can be measured by how close the value $Q$ of the partition found is to the value of $Q_{\max}$. 
Any solution provides a lower bound estimate of the value of $Q_{\max}$. Thus, the higher a solution's value of $Q$ is, the more accurate it and its estimate of $Q_{\max}$ is.

%It is of considerable interest and importance to develop an algorithm that robustly finds an accurate solution to this optimization problem that completes in polynomial time. The accuracy of a solution is determined by how high its value of $Q$ is.

A number of polynomial time complexity algorithms for finding a network partition that enables $Q_{\max}$ to be estimated have been proposed. Some are quite fast, such as random greedy agglomeration\cite{clauset2004finding,newman2004fast,ovelgonne2010cluster} and the Louvain method\cite{blondel2008fast}.
These algorithms, however, don't generally find very accurate solutions.
Far more accurate solutions can generally be found with spectral clustering algorithms\cite{newman2006finding,newman2006modularity} that iteratively bisect the set of nodes.  
The most accurate algorithm of this type~\cite{trevino2015fast} combines bi-sectioning based on the eigenvector of largest eigenvalue of the modularity matrix~\cite{newman2006finding}, tuning with generalized Kernighan–Lin refinements~\cite{kernighan_lin,sun2009improved}, and agglomeration.
Until recently this was the most accurate algorithm known.
Virtually all algorithms for maximizing modularity are partially stochastic, as they make random choices at intermediate steps among what are seemingly equivalent options at that point. These choices can affect the final partition, and, thus, different runs can produce different partitions. Because of this, to find the partition that provides the best estimate of the maximum modularity, algorithms are often run multiple times to produce an ensemble of partitions and the best of those partitions is chosen.

It has, however, recently been demonstrated that partitions with even more accurate estimates of $Q_{\max}$ can be obtained with a scheme that uses information contained within an ensemble of partitions generated with conventional algorithms.
This idea is known as ensemble learning. Its use distinguishes a new class of modularity maximizing algorithms~\cite{polikar2006ensemble,sagi2018ensemble}.
An ensemble learning scheme known as Iterative Core Group Graph Clustering (CGGCi)~\cite{ovelgonne2012ensemble} was the most accurate algorithm for finding the network partition that maximizes modularity in the 10\textsuperscript{th} DIMACS Implementation Challenge~\cite{dimacs}.
The CGGCi scheme starts with an ensemble of partitions obtained by using a conventional ``base algorithm'' and identifies ``core groups'' of nodes that are grouped together in the same community in every partition in the ensemble. 
It then transforms the original network into a weighted reduced network by collapsing each of these core groups into a single ``reduced'' node and summing all link weights between original nodes to assign weights to the links between the reduced nodes.
A base algorithm is then used to find an ensemble of partitions of the reduced network, and
that ensemble is used to find a new reduced network. This procedure is iterated until no further improvement in $Q$ is found. 
The best partition of the final reduced network is then mapped back onto the original network to identify the communities.

In this paper, we introduce a different ensemble learning scheme for network community detection. It uses an algorithmic paradigm we call Extremal Ensemble Learning (EEL).  
Our scheme, which we refer to as Reduced Network Extremal Ensemble Learning (RenEEL), starts with an ensemble of partitions obtained using a conventional base algorithm, and then iteratively updates the partitions in the ensemble until a consensus about which partition is best is reached within the ensemble.
To find the partitions used to update the ensemble efficiently, core groups of nodes are identified and used to form a reduced network that is partitioned using a base algorithm.
RenEEL then uses a partition of the reduced network to update the ensemble through extremal updating.
We will show that an algorithm using the RenEEL scheme improves the quality of community structure discovered, especially
for larger networks for which estimating the partition with $Q_{\max}$ becomes challenging. Testing our scheme on a wide range of real-world and synthetic benchmark networks, we show that it outperforms all other existing methods, consistently finding partitions with the highest values of $Q$ ever discovered.

\section*{Methods}
\subsection*{Community detection via modularity maximization}
Modularity $Q$ is a metric that quantifies the amount of modular structure there is in a given partition of a network's nodes into disjoint communities $P = \{ c_1, c_2, \ldots, c_r\}$, where $c_i$ is the $i$th community of nodes and $r$ is the number of communities. 
It is defined as~\cite{newman2004finding}
\begin{equation}
Q = \sum_{i}\left[\frac{m_i}{m}-\left(\frac{2m_i+e_i}{2m}\right)^2\right]
\label{Q}
\end{equation}
where the sum is over communities, $m_i$ and $e_i$ are respectively the number of internal and external links of community $c_i$, and $m$ is total number of links in the network. The first term in Eq.~\ref{Q} is the fraction of links inside communities, and the second term is the expected fraction if all links of the network were randomly placed.
For a weighted network, $m_i$, $e_i$ and $m$ are sums of link weights instead of numbers of links.
Modularity measures the deviation of the structure of a network partition from that expected in a random null model.
The {\it community structure of a network} corresponds to the partition $P$ of its nodes that maximizes $Q$. The number of communities in $P$ is free to vary.
The challenge of detecting the community structure of a network, therefore, is to find the partition with the maximum modularity $Q_{\max}$. 

\subsection*{Reduced networks}
\begin{figure}
\centering
\includegraphics{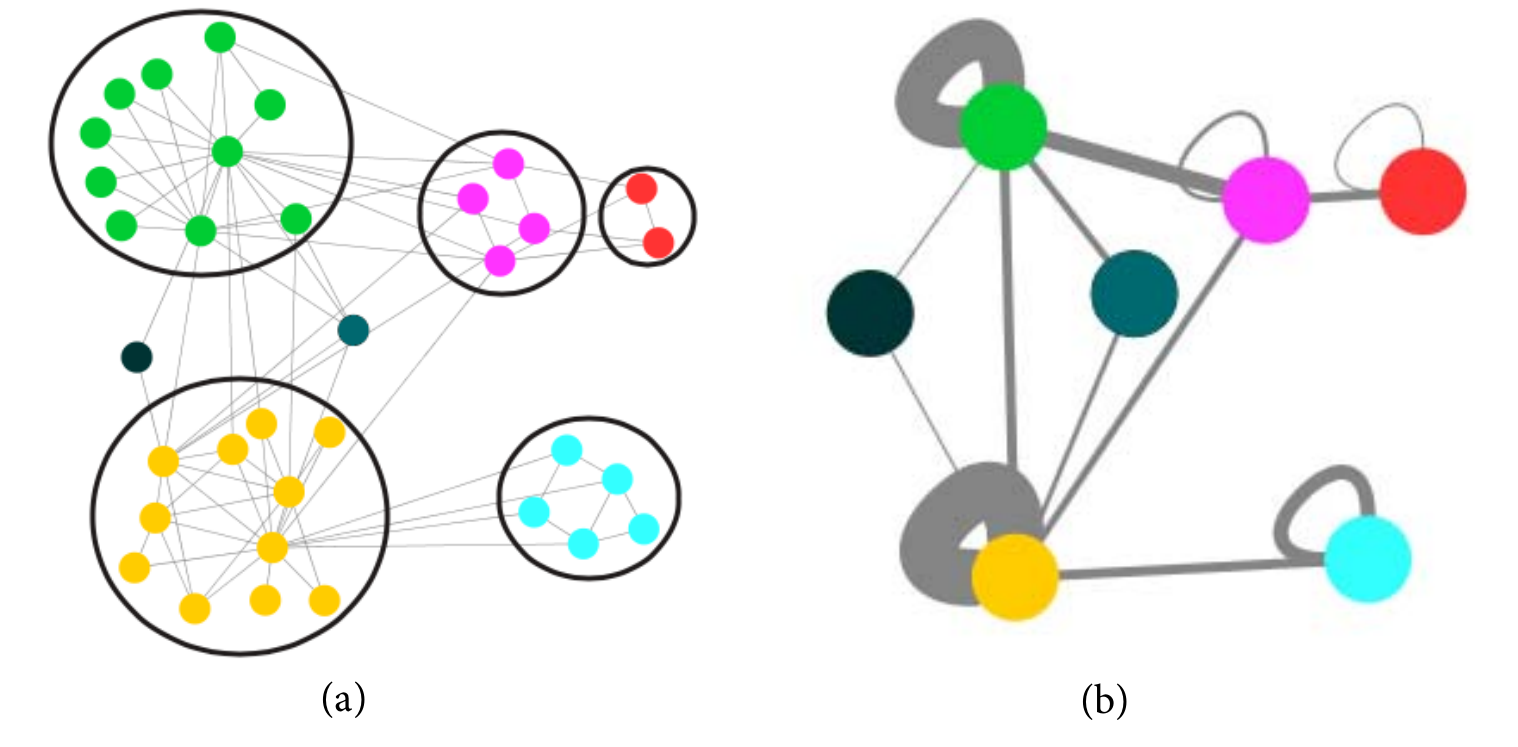}
\caption{\label{cg_network} {{\bf Construction of a reduced network.} (a) An example network showing seven core groups of nodes. The nodes of the same color belong to the same core group. The nodes inside each of the five circles are collapsed to single nodes in the reduced network, and the two isolated nodes also become nodes in the reduced network. (b) The reduced network after collapsing the core groups into single nodes. The nodes in the reduced network are colored according to the core group nodes in the original network and thickness of each link is proportional to its weight.}}
\end{figure}

To find a reduced network $G'$ starting from a network $G$ and an ensemble of partitions of it ${\cal P}$, we first identify the core groups in $G$. A \textit{core group} is a set of nodes that are found together in the same community in every partition in the ensemble. Any node that is not found in the same community with some other node in every partition in ${\cal P}$ is itself a core group.
$G'$ is then formed by collapsing core groups of nodes into single reduced nodes and combining their links to other nodes by summing their weights. An example of this is shown in Fig.~\ref{cg_network}. Each circle containing multiple nodes of $G$ that are colored the same in Fig.~\ref{cg_network}(a) denotes a core group. Two nodes that do not belong to any circle are shown in black and dark green. The core groups are collapsed to reduced nodes of the same color in the reduced network $G'$ shown in Fig.~\ref{cg_network}(b). The link weights in the reduced network are the sum of link weights between core groups in the original network. The weighted self-loops in $G'$ result from the total internal weights of the core groups in $G$.

\begin{figure}
\centering
\includegraphics{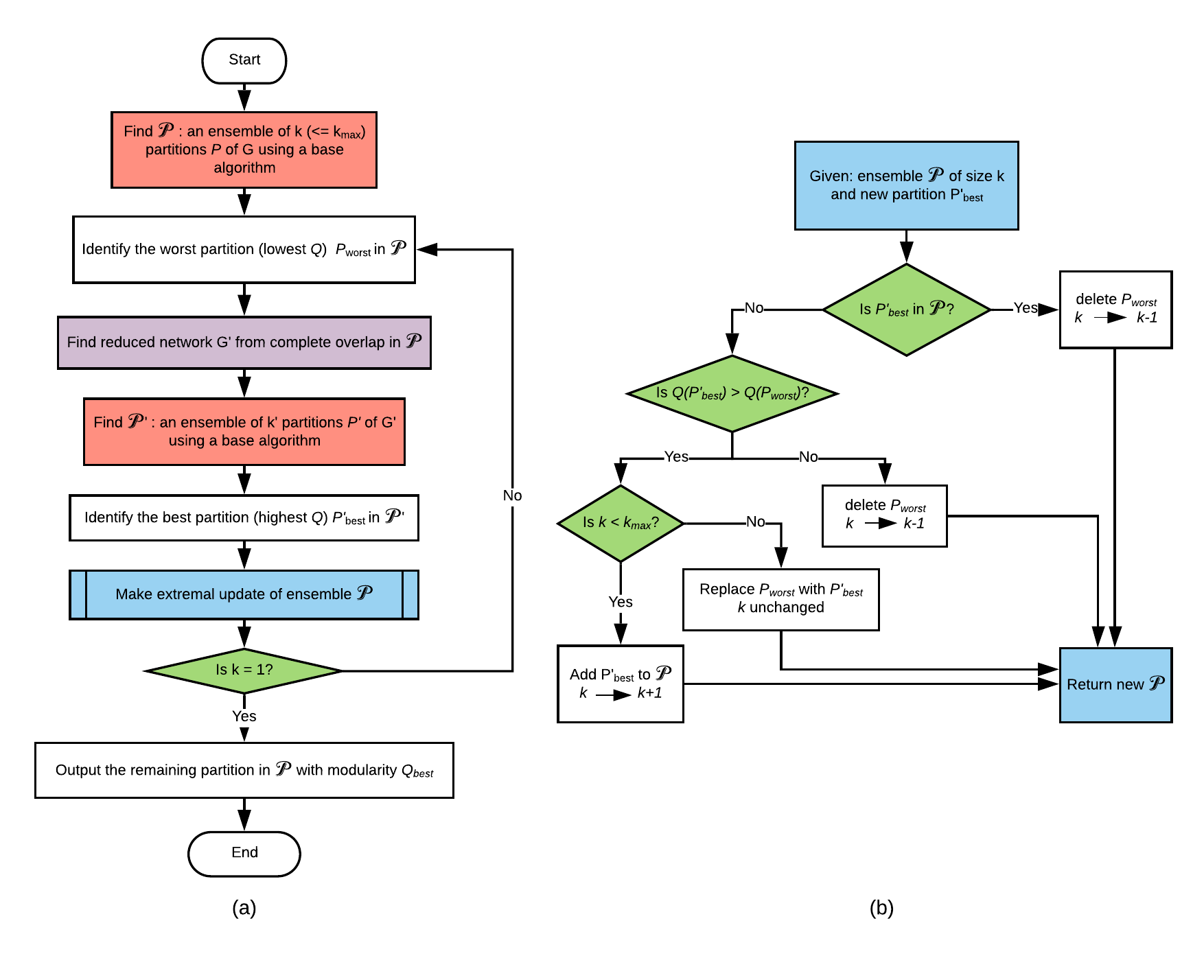}
\caption{\label{dcggc} {{\bf The RenEEL scheme.} (a) The steps of an efficient ensemble learning scheme to find the network partition that maximizes modularity $Q$ are shown in this flow chart. In the two steps shown in red a base algorithm is used to obtain an ensemble of partitions. The step shown in purple collapses the core groups to find the reduced network. The ensemble ${\cal P}$ gets updated with extremal criteria in the step shown in blue and is described in (b). The step shown in green guarantees algorithmic termination in a finite network. (b) The procedure of the extremal updating of ensemble ${\cal P}$.}}
\end{figure}

\subsection*{Reduced Network Extremal Ensemble Learning scheme}
The RenEEL scheme is summarized in the flowchart 
%\footnote{The flowcharts adhere to the ISO 5807:1985 standard (Information processing -- Documentation symbols and conventions for data, program and system flowcharts, program network charts and system resources charts).}
shown in Fig.~\ref{dcggc} and is described as follows.
First, an ensemble ${\cal P}$ of at most $k_{\max}$ partitions $P$ of the network $G$ is obtained from multiple runs of a base algorithm.
The base algorithm can be, for example, any of the conventional ones that have been developed to find a partition to estimate $Q_{\max}$. 
Alternatively, a set of base algorithms can be used to find ${\cal P}$.
The partitions in ${\cal P}$ are then ordered according to their modularity values, from the one with the largest value $P_{\rm best}$ to the one with the smallest value $P_{\rm worst}$.
Next, the core groups of nodes in the ensemble ${\cal P}$ are identified and used to construct the reduced network $G'$. An ensemble ${\cal P'}$ consisting of $k'$ partitions $P'$ of $G'$ is then obtained using a base algorithm. The base algorithm used for this step can either be the same as or different from the base algorithm used to find ${\cal P}$. The steps in which a base algorithm is used to find the ensembles ${\cal P}$ and ${\cal P'}$ are shown in red in Fig.~\ref{dcggc}(a).
The partition in ${\cal P'}$ with the largest modularity value $P'_{\rm best}$ is then identified and used to perform an extremal update of ensemble ${\cal P}$. This step is shown in blue in Fig.~\ref{dcggc}(a) and detailed in Fig.~\ref{dcggc}(b). 
If $Q(P'_{\rm best}) > Q(P_{\rm worst})$, then $P'_{\rm best}$ is expanded into a partition of $G$ and either used in place of $P_{worst}$ in ${\cal P}$ (if $k=k_{\max}$) or added to the ensemble $\cal P$ (if $k < k_{\max}$) as shown in Fig.~\ref{dcggc}(b).
In doing so ${\cal P}$ is enriched with a better quality partition. 
However, it is possible that at any iteration either $P'_{\rm best}$ is already contained in ${\cal P}$, or $Q(P'_{\rm best}) < Q(P_{\rm worst})$. In both cases, in order to move toward consensus within ${\cal P}$, its current size $k$ is reduced by 1 by deleting $P_{\rm worst}$ from it.
This procedure is repeated until there is only one partition left in the ensemble $\cal P$.
This consensus partition is the partition that has the largest modularity. It can be used to identify the communities of the network, and its modularity $Q_{\rm best}$ estimates $Q_{\max}$.

\subsection*{Computational complexity and practical implementation}
The most computationally complex and time consuming steps of the RenEEL scheme are those that use a base algorithm to find an ensemble of partitions. These steps are colored in red in the flowchart in Fig.~\ref{dcggc}. Assuming that the size of the ensembles ${\cal P}$ and ${\cal P'}$ are fixed,  the computational complexity of executing these steps is simply a fixed multiple of the computational complexity of the base algorithm used. The scaling of the computational complexity of base algorithms is typically between ${\cal O}(n^2)$ and ${\cal O}(n^3)$, where $n$ is the number of nodes in the network.
All other steps of the scheme have less complexity;
the steps of network reduction, colored purple in Fig.~\ref{dcggc}, and network expansion both have a computational complexity that scales as ${\cal O}(n^2)$, and the rest all have a computational complexity that is ${\cal O}(1)$. Thus, since each iteration of the scheme has only one step that uses the base algorithm a fixed number of times, each iteration has a computational complexity that scales the same as that of the base algorithm used. 
As the scheme progresses, however, the size of the reduced network monotonically decreases, significantly increasing the speed of later iterations.

A RenEEL algorithm applied to a finite network is sure to complete since new partitions are added to the ensemble ${\cal P}$ only if they have a modularity that is greater than $Q(P_{\rm worst})$ and the size of ${\cal P}$ is bounded.  
However, it is difficult to determine the precise scaling of number of iterations required in general for an algorithm implementing the scheme to complete, as it depends on the structure of the specific network under consideration. 
For the networks we analyzed, 
the number of iterations required was approximately proportional to $k_{\max}$.
Thus, we find empirically that the overall complexity of a RenEEL algorithm scales roughly as the base algorithm times $k'$ times $k_{\max}$.

The base algorithm used to obtain the results presented in this paper is a randomized greedy agglomerative hierarchical clustering algorithm~\cite{ovelgonne2010cluster}.
It is commonly used to find the community structure in complex networks~\cite{ovelgonne2012ensemble} and has an expected time complexity that scales as ${\cal O}(m \; \ln n)$~\cite{ovelgonne2010cluster}, where $m$ is the number of links in the network. There can be, at most, ${\cal O}(n^2)$ links. 
The overall complexity of the algorithm used here thus scales approximately as ${\cal O}(k_{\max}k' n^2 \; \ln n)$.
The particular choice of parameters $k_{max}$ and $k'$ is important for the quality of community structure as well as the computational time. In general, higher $k'$ and $k_{max}$ yield higher $Q_{\rm best}$. 

\subsection*{Co-clustering analysis}
\begin{figure}
\centering
\includegraphics{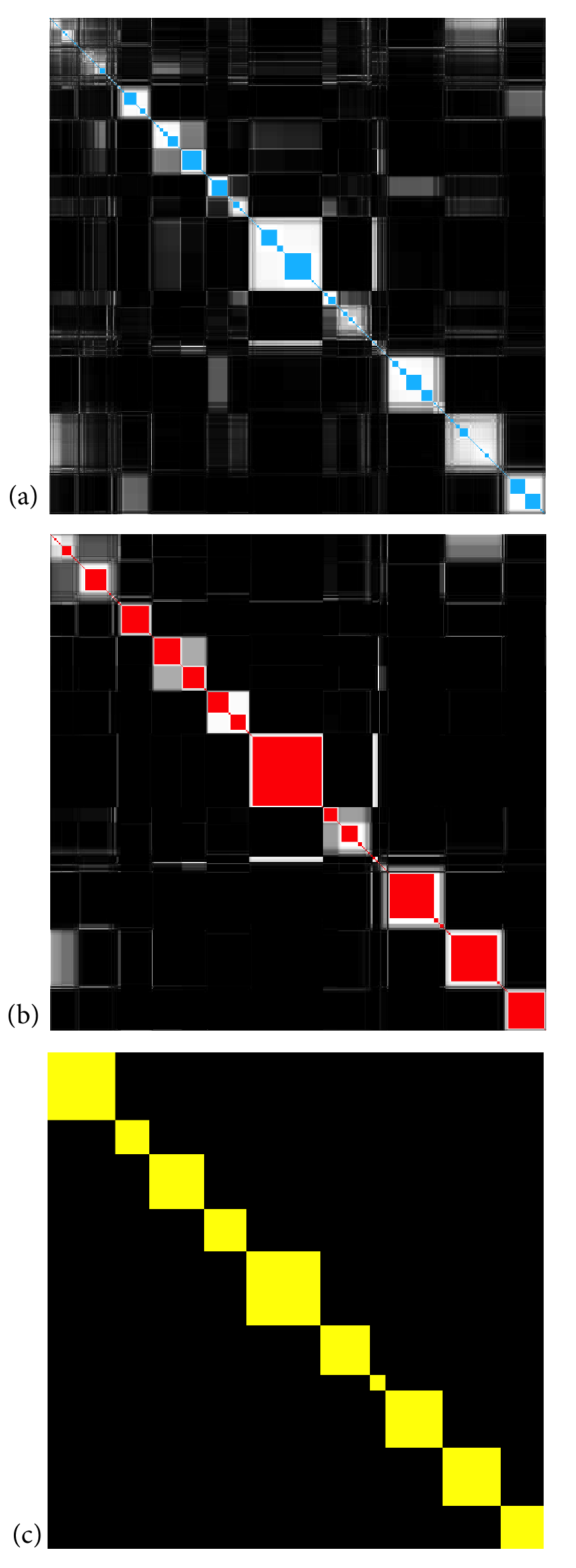}
\caption{\label{g123} {\bf Ordered co-clustering matrix with core groups.} Co-clustering matrix after the nodes have been reordered by simulated annealing. (a) after the first iteration (b) at the intermediate stage (c) at completion. The intensity of white in each pixel is proportional to the co-clustering frequency of the corresponding pair of nodes, except when the pair of nodes are always grouped together and, thus, belong to the same core group. In that case the pixel is colored blue in (a), red in (b), and yellow in (c).}
\end{figure}

\begin{figure}
\centering
\includegraphics{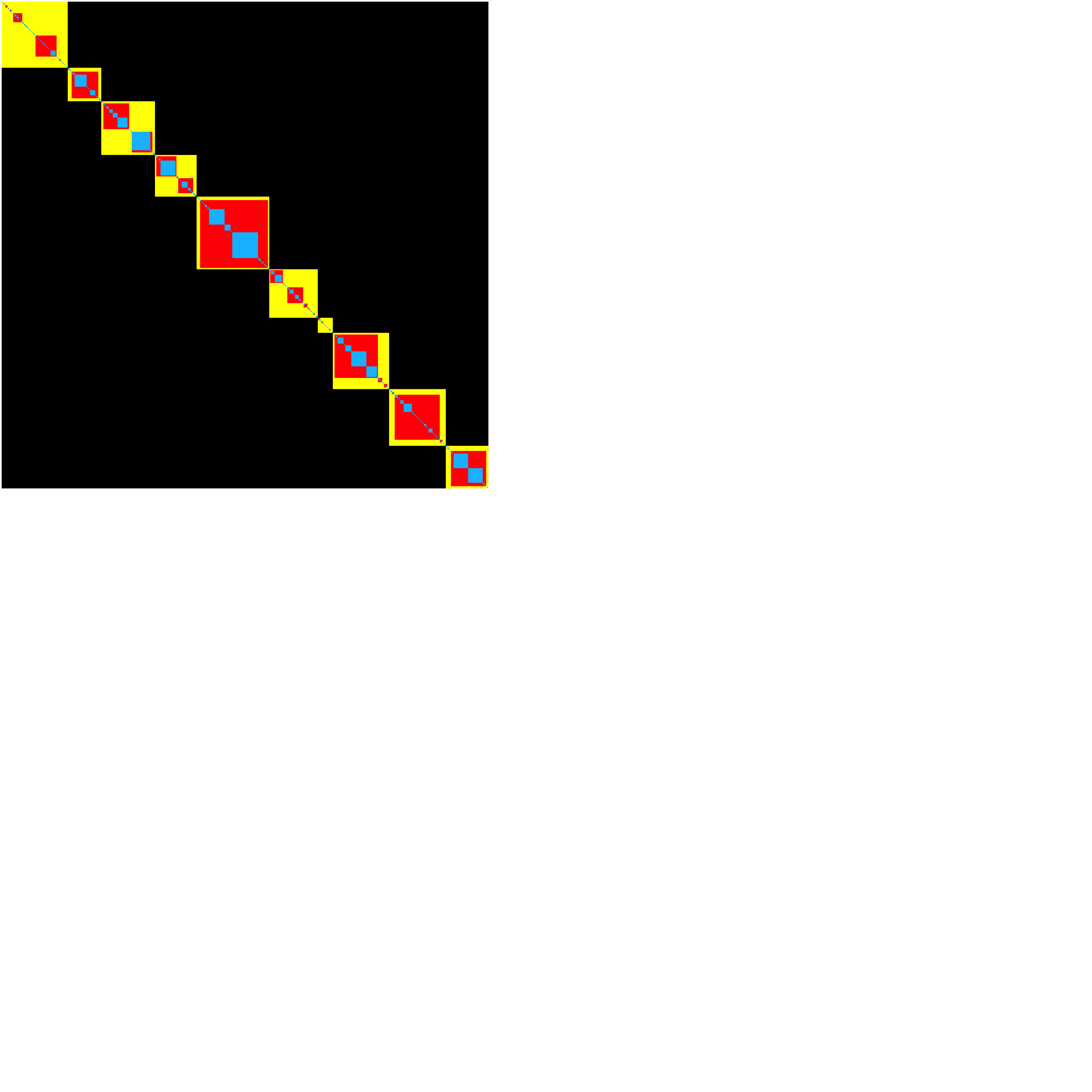}
\caption{\label{all_core_groups} {\bf Growth of core groups.} Colors blue, red, and yellow represent the core groups after the first iteration, at an intermediate stage, and at the end when the core groups have reached a stable state, respectively. The core groups can only grow. The process is agglomerative.}
\end{figure}

In order to visualize the evolution of the clustering results in the RenEEL scheme, co-clustering matrices at various stages of the scheme are shown in Fig.~\ref{g123}. 
In Fig.~\ref{all_core_groups} the results of the core group co-clustering at the different stages are combined to show their evolution.
%In that figure and in Fig.~\ref{all_core_groups}, which combines results of the core group co-clustering at the different stages, we use simulated annealing~\cite{kirkpatrick1983optimization} to order the nodes of the network.
A co-clustering matrix $S$ is a matrix whose elements $s_{ij}$ are defined as the fraction of times node $i$ and node $j$ are in the same community in an ensemble of partitions ${\cal P}$. 
The order of the nodes in Figs.~\ref{g123} and \ref{all_core_groups} was determined using simulated annealing to optimize the block-diagonal structure of the matrices.
Starting from a random ordering of the nodes, their order was rearranged to minimize a cost function, or ``Hamiltonian'', that is a function of minimum distance of matrix elements ($i,j$) from the diagonal $d_{ij}$ assuming periodic boundary conditions on the order:
\begin{eqnarray}
H = \sum_{i < j} s_{ij}\;d_{ij}^{\alpha}, 
\end{eqnarray}
where $\alpha$ is an arbitrary factor that controls the non-linear dependence of $H$ on $d_{ij}$.
The results in Figs.~\ref{g123} and \ref{all_core_groups} were obtained using $\alpha=3$.
Simulated annealing seeks to find the order of nodes that minimizes $H$. 
For the Monte Carlo updates in our simulated annealing, Metropolis rates~\cite{kirkpatrick1983optimization} with Boltzmann factor $e^{-(\Delta H)/T}$ were used. 
Starting from a relatively high temperature where the order of the nodes is random, the temperature was systematically lowered each Monte Carlo step until the node order stabilized. 

To get the three co-clustering matrices shown in Fig.~\ref{g123}, which respectively show results
at the initial, intermediate, and final stages of the RenEEL scheme, 
the following procedure was used in the simulated annealing Monte Carlo.   
First nodes were reordered by considering swaps of random pairs of nodes so as to minimize $H$ in the final stage co-clustering matrix.
Then, swaps of pairs of final stage core groups and swaps of pairs of nodes within the final stage core groups were considered to 
minimize $H$ in the intermediate stage co-clustering matrix.
Finally, swaps of pairs of final stage core groups, swaps of pairs of intermediate stage core groups within a final stage core group, and swaps of pairs of nodes within an intermediate stage core group were considered to 
minimize $H$ in the initial stage co-clustering matrix. 
The order of nodes that resulted is used in all three co-clustering matrices in Fig.~\ref{g123} and in Fig.~\ref{all_core_groups}.

\subsection*{Benchmark networks used for comparison}
To test the effectiveness of our methods of community detection we studied a set of networks. All of these networks were used in the 10\textsuperscript{th} DIMACS challenge.~\cite{dimacs} The networks are unweighted and undirected. They also have no self-loops. They may be connected or disconnected.  The networks we studied are listed and described in Table~\ref{table1}. These networks have been compiled from various sources and cover a wide range of sizes, functions and other characteristics. Hence, they are often used as benchmarks for testing community detection methods. The lists of links defining the Email, Jazz, PGPgc, Metabolic networks were downloaded from Ref.~[\citeonline{arenas_data}]. For Adjnoun, Polblog, Netscience, Power, Astro-ph, As-22july06, Cond-mat-2005, they were downloaded from Ref.~[\citeonline{newman_data}]. For Memplus, it was downloaded from Ref.~[\citeonline{memplus_data}]. For Smallworld and CAIDARouterLevel, they were downloaded from Ref.~[\citeonline{other_networks}].

%% list of networks -- 
%#nodes links are listed in table[]

\section*{Results}
\subsection*{Evolution of core groups}
The essence of how the RenEEL scheme works and why it is efficient can be seen by the evolution of the co-clustering of the nodes across the ensemble ${\cal P}$. Fig.~\ref{g123} shows the co-clustering results during a typical realization of the scheme on the Email network~\cite{guimera2003self} (see Table~\ref{table1}) at the initial, intermediate and final stages. In the three sub-figures, the intensity with which a pixel $(i,j)$ is colored white corresponds to the frequency that nodes $i$ and $j$ are in the same community in the member partitions of ${\cal P}$. The pixels colored blue, red, and yellow indicate that the nodes are in the same community in all member partitions. The nodes in the blue, red, and yellow blocks on the diagonal are the core groups that are used to form the reduced network. Nodes are listed in the same order in each of the three sub-figures. Fig.~\ref{all_core_groups} shows the evolution of just the core groups in the same realization.      

The Email network has $n=1133$ nodes. Initially, as shown in Fig.~\ref{g123}(a), there are 446 core groups, most of which contain only one or two nodes. After 100 iterations of the scheme, as shown in Fig.~\ref{g123}(b), the number of core groups is reduced to 192. Finally, in the stable state, after about 300 iterations of the scheme, only 10 core groups remain, as shown in Fig.~\ref{g123}(c). This reduction, from the original network of 1133 nodes to a reduced network of 10 nodes, is a tremendous simplification and greatly improves the overall speed of network clustering.

Within a network $G$ it is generally ``easy'' to determine that certain groups of nodes should be clustered together. All partitions group them together. These are the core groups of nodes. The hard work in finding the optimal partition is to determine whether nodes that are grouped together in only some of the partitions should indeed be in the same community, that is, to determine whether or not core groups should combine.
This is precisely what RenEEL focuses on.
The formation and evolution of core groups in RenEEL is an agglomerative process~\cite{Rokach2005}.  
Once a core group is formed, RenEEL never subsequently divides it.
As the scheme progresses, core groups grow and merge with each other and the number of core groups monotonically decreases. 

\begin{table}
\begin{center}
\caption{{\bf Benchmark networks}. A list of empirical and synthetic networks frequently used for benchmarking modularity optimization methods.}
 \begin{tabular}{||c c c||} 
 \hline
 {\bf Network} & {\bf Node description} & {\bf Link description} \\ [0.5ex] 
 \hline\hline
 Adjnoun~\cite{newman2006finding} & the most commonly occurring & pair of words that occur \\
 & adjectives and nouns in the novel    & in adjacent position \\
 & "David Copperfield" by Charles Dickens & in the text of the book \\ \hline

 Jazz~\cite{gleiser2003community} & musician & collaboration \\ \hline
 
Metabolic~\cite{duch2005community,jeong2000large,overbeek2000wit} & metabolites (e.g., proteins) & interaction between them \\
 & (in C. elegans) & \\ \hline

Email~\cite{guimera2003self} & members & email interchanges \\ \hline 
Polblog~\cite{adamic2005political} & weblogs on US politics & hyperlink \\ \hline

Netscience~\cite{newman2006finding} & scientists working on & coauthorship \\
 & network theory and experiment & \\ \hline

Power~\cite{watts1998collective} &  either a generator, & power supply line \\
 & a transformator or a substation & \\ \hline

PGPgc~\cite{boguna2004models} & users of the & interaction \\
 & Pretty Good Privacy (PGP) algorithm & \\ \hline

Astro-ph~\cite{newman2001structure} & scientists & coauthorship in \\ 
 & & preprints on the Astrophysics \\ 
 & & E-Print Archive between\\
 & & Jan 1, 1995 and December 31, 1999.\\\hline

Memplus~\cite{davis2011university} & memory circuit elements & connections \\ \hline

As-22july06~\cite{newman_data} & autonomous systems & data connection \\ \hline

Cond-mat-2005~\cite{newman2001structure} & scientists & coauthorship in \\ 
 & & preprints on the Condensed \\ 
 & & Matter E-Print Archive between\\
 & & Jan 1, 1995 and March 31, 2005.\\\hline
 
 Smallworld~\cite{watts1998collective} & synthetic & synthetic \\ \hline
 
 CAIDARouterLevel~\cite{CAIDA} & routers & links \\ \hline

 % \\ [1ex] 
%\hline

\end{tabular}
\label{table1}
\end{center}
\end{table}

\subsection*{Evolution of the ensemble ${\cal P}$}
A defining characteristic of RenEEL is that the ensemble of partitions ${\cal P}$ evolves as the scheme progresses. The ensemble ``learns'' what the partition with $Q_{\rm best}$ is by using extremal updating
to incorporate new partitions, replace existing ones with higher quality ones, or remove low quality partitions.
The new partitions are partitions of the reduced network $G'$. 
They are used in RenEEL to improve the quality of ${\cal P}$ at every iteration of the scheme until a consensus is reached about what the optimal partition is. 

\begin{figure}
\centering
\includegraphics{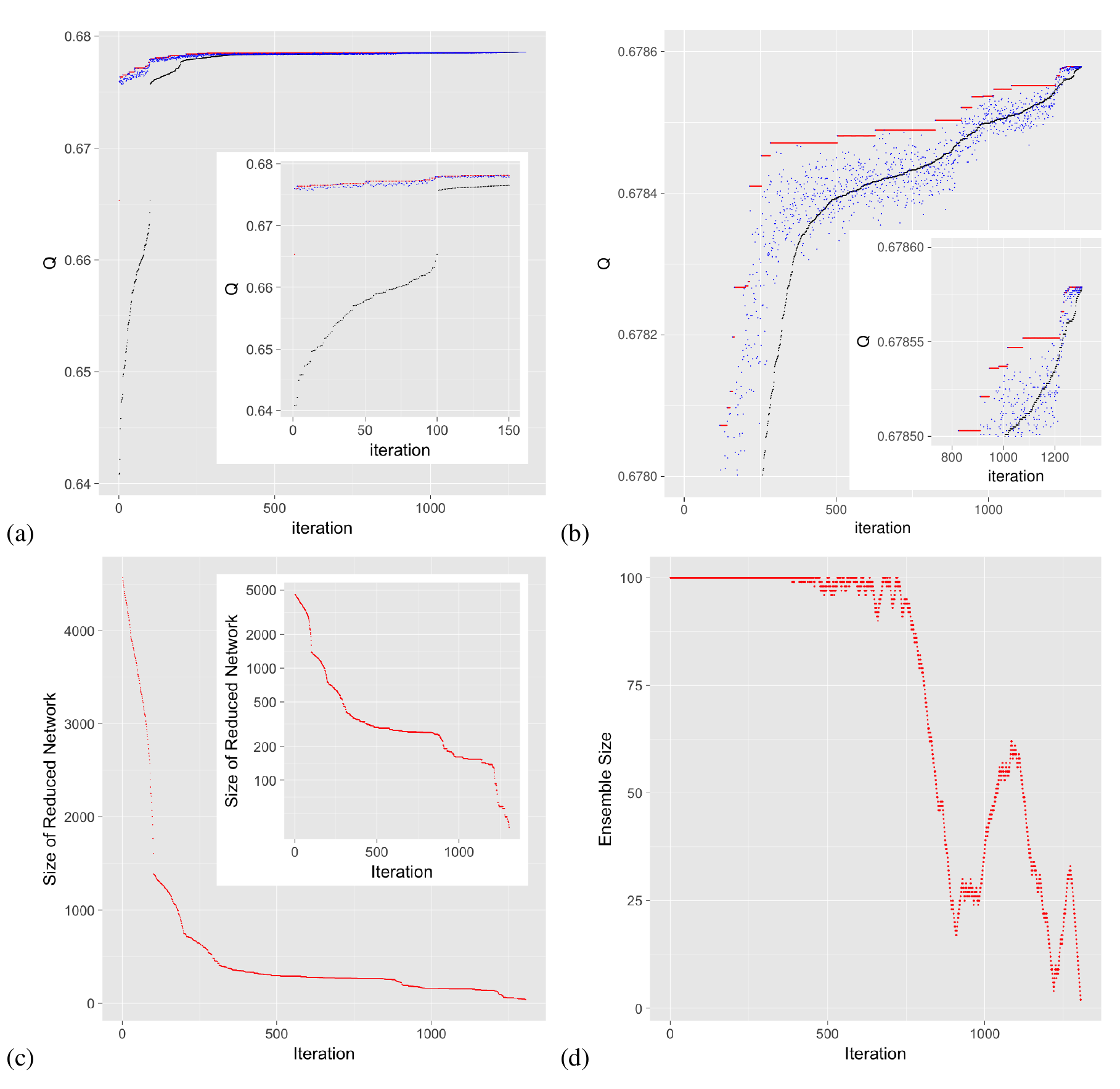}
\caption{\label{eel_iter} {\bf Evolution of the ensemble of partitions ${\cal P}$ for a typical run of RenEEL.} (a) Modularity $Q$ of partitions $P_{\rm best}$, $P'_{\rm best}$ and $P_{\rm worst}$ at each iteration of the scheme is shown in red, blue, and black, respectively. The inset is an enlargement of the results for the first 150 iterations. (b) Same results as in (a), but showing only the upper portion of the plot. The inset shows an enlargement of the upper-right corner of the plot. (c) Evolution of the size of the reduced network. The inset shows the same plot on a logarithmic scale. (d) Evolution of the size of the ensemble ${\cal P}$.}
\end{figure}

A typical way that ${\cal P}$ evolves as the scheme progresses can be seen with the results shown in Fig.~\ref{eel_iter} from an example run of RenEEL that partitions the As-22july06 network~\cite{newman_data}. (See table~\ref{table1}.) In this example run, $k_{\max}=100$ and $k'=20$.
Fig.~\ref{eel_iter}(a) and (b) show the modularity value $Q$ of $P_{\rm best}$ the best partition in ${\cal P}$ (red dots), of $P_{\rm worst}$ the worst partition in ${\cal P}$ (black dots), and of $P'_{\rm best}$ the new partition of $G'$ considered for the enrichment of ${\cal P}$ (blue dots) as a function of the number of iterations.
The main panel of Fig.~\ref{eel_iter}(a) shows the full results of the scheme, from start to finish.
An enlarged view of the results for the initial 150 iterations is shown in the inset of Fig.~\ref{eel_iter}(a).   
The main panel of Fig.~\ref{eel_iter}(b) shows an enlarged view of the vertical $Q$ axis near the final result of the entire scheme.
An enlarged view of both axes at the end stages of the scheme is shown in the inset of Fig.~\ref{eel_iter}(b).   
Fig.~\ref{eel_iter}(c) shows the size of the reduced network, or equivalently the number of core groups, as a function of the number of iterations. The main panel of Fig.~\ref{eel_iter}(c) shows the results on linear axis scales, and the inset shows the same results on log scales.
Fig.~\ref{eel_iter}(d) shows the ensemble size $k$ as function of the number of iterations.

In the example run, as can be seen from the inset of Fig.~\ref{eel_iter}(a), for the first 100 iterations the modularity of the new partitions $Q(P'_{\rm best})$ are all significantly better than that of the worst in the ensemble $Q(P_{\rm worst})$. In fact, all the first 100 new partitions generated by RenEEL are better than every one the 100 original ones in ${\cal P}$ generated by the base algorithm. (The number of partitions in ${\cal P}$ initially is $k_{\max}=100$.) So, for the first $k_{\max}$ iterations RenEEL systematically replaced each of the original partitions. There is large increase in $Q(P_{\rm worst})$ at iteration 100. Although it's difficult to see in the figure, there are other similar, significant increases in $Q(P_{\rm worst})$ at iterations 200 and 300, indicating that RenEEL also replaces its first and second 100 new partitions with entirely new sets in the second and third 100 iterations, respectively. After the first 300 iterations, the quality of the new partitions starts to become comparable to the existing partitions.
Throughout the process, the $Q(P_{\rm best})$ intermittently raises when a new best partition is discovered.

Fig.~\ref{eel_iter}(c) shows that the size of the reduced network keeps decreasing as the scheme progresses. It initially decreases exponentially, then there is what appears to be a power-law decay from iteration 100 to iteration 1000 (see inset of Fig.~\ref{eel_iter}(c)), followed by a sharp, perhaps exponential, decay in the final iterations of the scheme. The original size of this network, $n = 22963$, is reduced to $38$ core groups at the termination step. 
The size of the ensemble, shown in Fig.~\ref{eel_iter}(d), 
varies when new partitions are discovered and added to ${\cal P}$ or when low quality partitions are deleted as the scheme drives ${\cal P}$ toward consensus. 
The plot shows that as the ensemble learns, its size grows and shrinks multiple times before its size falls to unity and the scheme terminates. 
There are two main periods in which the size of the ensemble grows, one beginning at about iteration 900 and the other at about iteration 1200. During these periods the value of $Q(P_{\rm best})$ increases quickly, as can be seen in the main panel and inset of Fig.~\ref{eel_iter}(b). These are periods when the ensemble ${\cal P}$ has made a ''breakthrough'' by discovering a new set of high quality partitions.    
The example run ends with a consensus choice that a partition with modularity $Q_{\rm best} = 0.678579$ is the one that maximizes modularity for this network, a value higher than that any previously reported partition. (See Table~\ref{table2}.)

\subsection*{Distribution of results for $Q_{\rm best}$}
\begin{figure}
\centering
\includegraphics{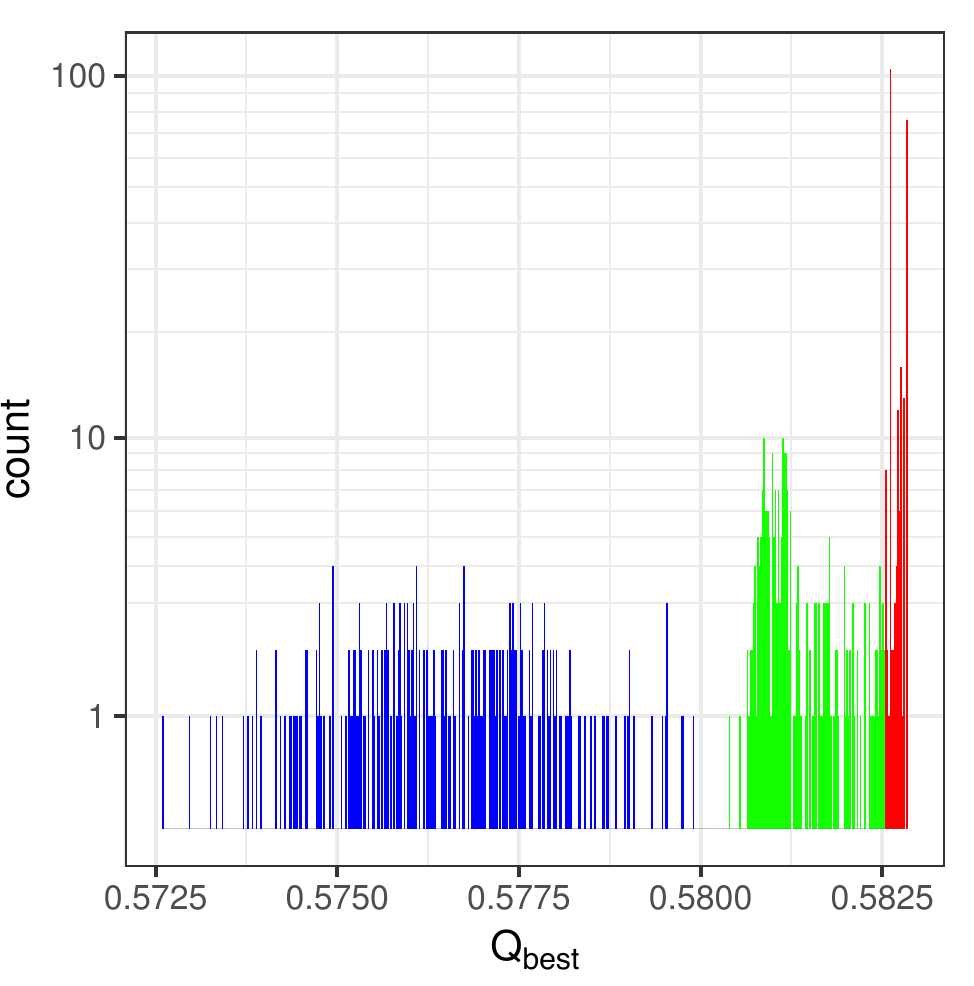}
\caption{\label{q_dist} {\bf Distribution of $Q_{\rm best}$ obtained by various methods.} Frequency plot of $Q_{\rm best}$ for the Email network obtained by multiple realizations of three different methods. Blue corresponds to a naive ensemble analysis scheme, green corresponds to CGGCi scheme, and red corresponds to RenEEL scheme. The y-axis has a logarithmic scale. In this particular example, there is no overlap between the distributions from the different methods.}
\end{figure}

Since virtually all conventional algorithms are stochastic, ensemble learning schemes that use them as base algorithms will also be stochastic. Thus, a range of results for $Q_{\rm best}$ are possible with each realization of virtually all methods of modularity maximization.
As an example, Fig.~\ref{q_dist} shows the distribution of $Q_{\rm best}$ that three different methods of community detection produce for the Email network. Results from 250 realizations for each method are shown.
Results from the RenEEL, CGGCi ensemble learning schemes, and naive ensemble analyses are shown in red, green, and blue, respectively.
The results for all three of these schemes were obtained using a randomized greedy algorithm as the base algorithm and
an ensemble size of $k_{\max}=100$. 
Each of the blue data points were obtained by running the algorithm 100 times and choosing the largest value from those runs.
The distributions from the three different methods are all non-overlapping, with the RenEEL results having the largest values, followed those of CGGCi and then those of the naive ensemble analyses with the conventional algorithm.
The distribution of $Q_{\rm best}$ for RenEEL is also narrower than those of the other two schemes,
which suggests that the results from RenEEL are close to the value of $Q_{\max}$ for the network.

\subsection*{Application to benchmark networks}
To test the accuracy of the RenEEL scheme, we applied it to the benchmark networks listed in Table~\ref{table1}. In Table~\ref{table2}, the maximum modularity value $Q_{\rm best}$ found for these networks by RenEEL is compared to the best previously published values. Many of these values were the best result in the 10\textsuperscript{th} DIMACS challenge~\cite{dimacs_results}.
To be consistent, all realizations had $k_{\max}=100$ and $k'=20$ and used the randomized greedy algorithm as a base network.
100 different realizations of RenEEL were run on the smaller networks, up to and including the 
Netscience network, and 5 were run on the larger networks. 
For the smaller networks the value of $Q_{\rm best}$ reported in table was consistently obtained. For the larger networks a range of results were obtained and the largest one is listed.
As the table shows, the partitions found by RenEEL have a value of $Q_{\rm best}$ that is higher than or equivalent to the best previously reported value for every benchmark network. 
The difference between $Q_{\rm best}$ found by RenEEL and the previous best values increases with network size. 
This is due to the fact that for small networks it is generally easier to find the Modularity maximizing partition, but the task becomes more challenging for larger networks.

Our results are significant for every network studied.
For the smallest networks, our best partition has the same modularity as that of the previous best result. This is presumably because we find the true best partition that other algorithms have also found. For larger networks, however, our results are better than any previously reported result.  For some medium size networks, our value of $Q_{\rm best}$ may be only slightly better than the previous best, but, in these cases, finding any new better result is remarkable and mathematically noteworthy. Furthermore, for these networks, we may be discovering the true best partition. For larger networks, our accuracy improvement is substantial. 

Perhaps a better way of quantifying the mathematical significance of our results would be, if one knew what the value of true best Modularity $Q_{\max}$ is, to consider results for $1/\Delta Q$, where $\Delta Q \equiv Q_{\max}-Q$, instead of the results for $Q$. Unfortunately, that’s not possible as the value of $Q_{\max}$ for most networks is not known. If we could though, it would be clear that our results are indeed highly significant, for every network studied.   

\begin{table}
\begin{center}
\caption{{\bf Comparison of results using RenEEL to the previous best results for benchmark networks.} Maximum modularity $Q_{\rm best}$ obtained by the RenEEL scheme compared to the previous best reported values.}
 \begin{tabular}{||c c c c c||} 
 \hline
 {\bf Network} & {\bf Nodes} & {\bf Links} & {\bf RenEEL result} & {\bf Previous best} \\ [0.5ex] 
 \hline\hline
Adjnoun & 112 & 425 & 0.313367 & 0.313367~\cite{Aloise2012ModularityMI} \\ \hline
Jazz & 198 & 2742 & 0.445144 & 0.445144~\cite{Aloise2012ModularityMI} \\ \hline
Metabolic & 453 & 2025 & 0.453248 & 0.453248~\cite{Aloise2012ModularityMI} \\ \hline
Email & 1133 & 5451 & 0.582829 & 0.582829~\cite{Aloise2012ModularityMI} \\ \hline 
Polblog & 1490 & 16715 & 0.427105 &	0.427105~\cite{Aloise2012ModularityMI} \\ \hline
Netscience & 1589 &	2742 &	0.959900 &	0.959900~\cite{ovelgonne2012ensemble}\\ \hline
Power & 4941 &	6594 &	0.940938 &	0.940851~\cite{Aloise2012ModularityMI} \\ \hline
PGPgc & 10680 &	24316 &	0.886853 &	0.886564~\cite{dimacs_results} \\ \hline
Astro-ph & 16706 &	121251 &	0.745614 &	0.744621~\cite{Aloise2012ModularityMI} \\ \hline
Memplus & 17758 &	54196 &	0.700591 &	0.700473~\cite{dimacs_results} \\ \hline
As-22july06 & 22963 & 	48436 &	0.678579 &	0.678360~\cite{ovelgonne2012ensemble} \\ \hline
Cond-mat-2005 & 40421 &	175693 &	0.748187 &	0.746445~\cite{ovelgonne2012ensemble} \\ \hline
Smallworld &  100000 & 499998 & 0.793175 & 0.793099~\cite{ovelgonne2012ensemble} \\ \hline
CAIDARouterLevel & 192244 & 609066 & 0.872086 & 0.872042~\cite{dimacs_results} \\ \hline

% \\ [1ex] 
%\hline

\end{tabular}
\label{table2}
\end{center}
\end{table}

\section*{Discussion}
Recent advances in Machine Learning and Artificial Intelligence have enabled progress to be made toward solving a range of difficult computational problems~\cite{mohammed2016machine}.
In this paper, we have introduced a powerful algorithmic paradigm for graph partitioning that we call Extremal Ensemble Learning (EEL). EEL is a form of Machine Learning. An EEL scheme creates an ensemble of partitions and then uses information within the ensemble to find new partitions that are used to update the ensemble using extremal criteria. Through the updating procedure, the ensemble learns how to form improved partitions, as it works toward a conclusion by achieving consensus among its member partitions about what the optimal partition is. 

The particular EEL scheme we have introduced, Reduced Network Extremal Ensemble Learning (RenEEL), uses information in the ensemble of partitions to create a reduced network that can be efficiently analyzed to find a new partition with which to update the ensemble. We have used RenEEL to find the partition that maximizes the modularity of networks. This is a difficult, NP-hard computational problem\cite{brandes2008modularity}. 
We have shown that an algorithm using the RenEEL scheme outperforms all existing modularity maximizing algorithms when analyzing a variety of commonly studied benchmark networks. For those networks it finds partitions with the largest modularity ever discovered. For the larger benchmark networks, the partitions that we 
discovered are novel.

Although we have only demonstrated the effectiveness of our algorithm for the well-known problem of finding the network partition that maximizes modularity, the EEL paradigm and the RenEEL scheme can be used to solve other network partitioning problems. For example, the algorithm we used can be straightforwardly adapted to optimize other metrics such as modularity density~\cite{chen2014}, or excess modularity density~\cite{chen2018network}. Work is underway to explore the effectiveness of RenEEL for solving those problems. Its potential effectiveness for finding the partition that maximizes excess modularity density may be especially important. Using excess modularity density largely mitigates the resolution limit problem in community detection by maximizing modularity~\cite{fortunato2007resolution}, making it a preferred metric for applications where the resolution limit is problematic, such as finding the community structure in gene regulatory networks~\cite{trevino2012robust,mentzen2008regulon}.

There is potential to improve upon our results using the RenEEL scheme. 
As previously discussed, any conventional algorithm can be used as the base algorithm of the scheme. There is also freedom to vary the size of the ensembles used in the scheme. Which base algorithm and what ensemble sizes are best to use depends on the network to be analyzed.
Using a high quality base algorithm though, such as the Iterative Spectral Bisectioning, Tuning and Agglomeration algorithm~\cite{trevino2015fast}, is likely to yield more accurate results for many of the networks studied.
There is also potential to improve the RenEEL scheme itself.
For instance, currently, a naive ensemble analysis of partitions of the reduced network is used to find a new partition with which to update the ensemble. Another method, such as a recursive use of the RenEEL scheme, may yield better results.
Also, currently, once the original ensemble of partitions is created, no new information is ever added to the system during the learning processes. It may be beneficial to occasionally use a new partition of the original network instead of the reduced network to update the ensemble. 
Work is in progress to explore if these ideas lead to improved results.

Finally, the principal reasons why the RenEEL scheme is both efficient and effective should be noted. 
Its efficiency stems from its use of an ensemble of partitions to form reduced networks. 
The smaller size of the reduced networks allows them to be partitioned much more quickly than the full network.
Also, because the scheme is so effective, highly accurate results can be obtained even if a fast, but low quality, base algorithm is used. This allows significantly larger networks to be analyzed than what would otherwise be possible.
The remarkable effectiveness of RenEEL, even relative to other Ensemble Learning schemes, is mainly due to its extremal updating of the ensemble of partitions. 
It is of course just one example of a scheme using the EEL paradigm. Its success, though, suggests that EEL is an algorithmic paradigm that will be useful for solving a variety of graph theoretic problems.

%\noindent LaTeX formats citations and references automatically using the bibliography records in your .bib file, which you can edit via the project menu. Use the cite command for an inline citation, e.g.  \cite{Hao:gidmaps:2014}.

%For data citations of datasets uploaded to e.g. \emph{figshare}, please use the \verb|howpublished| option in the bib entry to specify the platform and the link, as in the \verb|Hao:gidmaps:2014| example in the sample bibliography file.

\section*{Acknowledgements}
We thank Peter Grassberger and Eve S. Wurtele for fruitful discussions. This work was supported by the NSF through grants DMR-1507371 and IOS-1546858. Some of the computations in this work were done on the uHPC cluster at the University of Houston, acquired through NFS Award Number 1531814.
\section*{Author contributions statement}
JG, PS, and KEB conceived of the project. JG performed the simulations. JG, PS, and KEB analyzed the results and wrote the paper. All authors read and approved the manuscript.
\section*{Additional information}
\textbf{Competing interests}
The authors declare that they have no competing interests.
%To include, in this order: \textbf{Accession codes} (where applicable); \textbf{Competing interests} (mandatory statement). 

%The corresponding author is responsible for submitting a \href{http://www.nature.com/srep/policies/index.html#competing}{competing interests statement} on behalf of all authors of the paper. This statement must be included in the submitted article file.
\section*{Data Availability}
The data used in this study are publicly available from the sources that are cited in the main text.

\end{document}